\documentclass{ceab}   

\usepackage{epsfig}     
\usepackage{graphicx}   
\usepackage{url}
\usepackage{natbib}     
\usepackage[T1]{fontenc} 
\usepackage{babel}       
    
\def\runningtitle{Ru\v{z}djak et al.: ECRS2024 Proceedings}   
\def\articlenumber{7}
\def\ppages{1--4}

\begin{document}

\title{An improved method to search for flares from point sources of ultra-high-energy photons}

\author{
  J.~Stasielak$^1$,  N.~Borodai$^1$,
      D.~G\'ora$^1$,
      M.~ Niechciol $^2$
\vspace{2mm}\\
\it $^1$Nuclear Physics Polish Academy of Sciences, PL-31342 Krakow, Poland\\
\it $^2$Center for Particle Physics Siegen, University of Siegen, Siegen, Germany}

\maketitle

\begin{abstract}
Flares produced by certain classes of astrophysical objects may be sources of some ultra-high-energy particles, which, if they are photons, would group into clusters of events correlated in space and time. Identification of such clustering in cosmic-ray data would provide important evidence for possible existence of ultra-high-energy (UHE) photons and could potentially help identify their sources. We present an analysis method to search for space-time clustering of ultra-high-energy extensive air showers, namely the stacking method, which combines a time-clustering algorithm with an unbinned likelihood study. In addition, to enhance the capability to discriminate between signal (photon-initiated events) and background (hadron-initiated) events, we apply a photon tag. This involves using relevant probability distribution functions to classify each event as more likely to be either a photon or a hadron. We demonstrate that the stacking method can effectively distinguish between events initiated by photons and those initiated by hadrons (background). The number of photon events in a data sample, as well as the flare(s) duration can also be retrieved correctly. The stacking method with a photon tag requires only a few events to identify a photon flare. This method can be used to search for the cosmic ray sources and/or improve limits on the fluxes of UHE photons.
\end{abstract}


\section{Introduction}

Some fraction of ultra-high-energy cosmic rays (UHECRs) may originate from astrophysical flaring events such as AGN flares or gamma-ray bursts. This could potentially lead to the formation of clusters in the observational data collected by various cosmic-ray experiments, meaning the formation of groups of events in the data that are correlated both temporally and spatially. Space-time correlations of events (clustering) are most likely to occur for neutral particles, such as photons, that are not deflected by magnetic fields and therefore follow a straight path to Earth, pointing to their sources. Thus, the observation of clustering in the cosmic-ray data may indicate the existence of UHE photons, i.e. with energy larger than $10^{17}$ eV.

In this paper, we propose an improved search algorithm for 
space-time clustering in UHECR (air-shower) data,
a slight modification \citep{GORA2011201} of the commonly used unbinned likelihood method \citep{BRAUN2008299},  which can be applied to a time-dependent search for weak multiple flares from point sources of UHE photons. 
The proposed algorithm has already been applied to investigate correlations between the arrival directions of neutrinos detected by the IceCube Neutrino Observatory and the directions to their possible sources \citep{GORA2011201}.

We demonstrate that this improved approach, so called stacking method, is capable of effectively detecting time-space-correlated UHECR clusters. Moreover, we calculate its efficiency. This method may prove useful in identifying cosmic ray sources and in establishing stronger constraints on the fluxes of UHE neutral particles such as UHE photons. 

\section{Standard space-time clustering analysis}

The standard approach to search for 
space-time clustering in UHECR (air-shower) data
is based on the unbinned likelihood method \citep{BRAUN2008299} with an addition of search for the time correlation between events. In this method, finding a point source of photons in the sky (flaring event) means to locate an excess of space-time correlated events from a particular direction over the background (a cluster of events). The overall procedure for such a search is as follows. First, we take all consecutive multiplets in the data, i.e. doublets, triplets, quadruplets, and so on. Subsequently, for each multiplet $j$ we construct likelihood $\mathcal{L}$ and the corresponding test statistic $\rm{TS}$. Next, we maximize $\rm{TS}$ 
to obtain its maximum value $\rm{TS}_{\max}$.
As an outcome, we get a set of $\rm{TS}_{\max}$ values, one for each multiplet. We choose the most significant multiplet, i.e. the one with the largest $\rm{TS}_{\max}$. The flare duration $\Delta T$ is estimated to be the length of the time window of this most significant multiplet. 
The statistical significance of the results is assessed by comparing the value of $\rm{TS}_{\max}$ calculated for the most significant multiplet of the data with a distribution of maximum values of the test statistic 
derived from many randomly generated scrambled maps mimicking the original dataset. Details of the described method are given below.


A signal probability distribution function (PDF) of an event $i$ is given by
\begin{equation}
s_i=s_i^{\mathrm{space}}s_i^{\mathrm{time}} \rm{,}
\label{llh_sig_time}
\end{equation}
a combination of the contribution from space and time part. The spatial probability $s_i^{\mathrm{space}}$ is a Gaussian function
\begin{equation}
s_i^{\rm{space}} = \frac{1}{2\pi \sigma^2} \exp \left(-\frac{\left|\vec r_i - \vec r_s \right|^2}{2\sigma_i^2}\right) \rm{,} 
\end{equation}
where the angular reconstruction uncertainty of considered event and its direction are given by $\sigma_i$ and $\vec r_i$, respectively. The source direction is denoted by $\vec r_s$. 
For each time window of a multiplet tested $\Delta t_j = t^{\rm{max}}_j - t^{\rm{min}}_j$, a temporal PDF is given by
\begin{equation}
 s_i^{time}=\frac{H\left ( t_{j}^{max}-t_i \right )H\left ( t_i-t_{j}^{min} \right )}{\Delta t_{j}},
 \label{sig_time}
\end{equation}
where $H$ is the Heaviside step function and $t_i$ is the arrival time of event. Note, that by using this definition, we count only events that fall within
the considered time window  $\Delta t_j$, therefore $s_i=0$ outside of it.

Similar to the signal PDF, the  background  PDF also contains 
space and time part, so it can be written as
\begin{equation}
b_i=b_i^{\mathrm{space}}b_i^{\mathrm{time}}.
\label{llh_bg_time}
\end{equation}
To define the spatial and temporal background PDFs we use the total solid angle $\Delta \Omega$ subtended by the considered part of the sky, and the length of the entire data taking period, or more precisely, the uptime $\Delta T_{\mathrm{data}}$. The resulting PDFs are $b_i^{\mathrm{space}} = 1 / \Delta \Omega$ and $b_i^{\mathrm{time}}=1 / \Delta T_{\mathrm{data}} $.

%
%

Both the signal and background PDFs are combined to evaluate the likelihood over all observed events (numbered by index $i$) such that
\begin{equation}
\mathcal{L}(n, \Delta t_j,\vec r_s) = \prod_{i=1}^{N} \Bigg(\frac{n}{N}s_{i} + (1 - \frac{n}{N}){b}_i\Bigg) \rm{,} \label{aa}
\end{equation}
where $n$ is the assumed number of signal events present in the cluster and $N$ is the number of all events in the considered data sample. Following this equation, a likelihood representing absence of signal events $\mathcal{L}(0, \Delta t_j, \overrightarrow{r_s})$, i.e. containing only the background, can also be calculated. Then by comparing these two likelihoods, a test statistic evaluating the significance of a given multiplet, can be constructed: 
\begin{equation}
\rm{TS}(n) = -2  \cdot log\Bigg[\frac{\mathcal{L}\left( 0, \Delta t_j,{\vec r_s} \right)}{\mathcal{L}(n, \Delta t_j, \vec r_s)}\Bigg].
\label{bb}
\end{equation}
The $\rm{TS}(n)$ should be then maximized against $n$ to obtain an estimate of the number of signal events, denoted by $n_s$. Note that by construction $n_s$ is a real number and not necessarily an integer, because TS is maximized over the entire range of real numbers rather than over a discrete set of integers. However, we can expect integers to occur more frequently, thus forming multi-peak distributions of $n_s$. The TS distribution obtained from many scrambled maps is used to evaluate the significance level of the results.

\label{first}

\section{The stacking method}

We propose an improved method of the space-time clustering analysis \citep{GORA2011201}, which involves a slight modification of the standard likelihood approach \citep{BRAUN2008299} described in the previous section. This improved method is based on doublet stacking analysis, offering significantly faster performance than the standard method, and making it sensitive to multiple weak flares of arbitrary shapes. 
The method consists of 3 steps. 

\begin{figure*}[t!]
\vspace{-0.8cm}
\begin{center}
\includegraphics[width=1\textwidth]{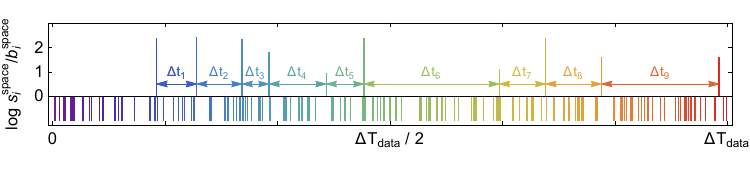}
\end{center}
\vspace{-1cm}
\caption{All events collected over time period $\Delta T_{\rm{data}}$ projected onto the timeline. Each line represents a single event. Time of events is color coded, blue means an earlier time, and red later time. Lines extending above black horizontal line, defined by $\log ( s_i^{\rm{space}}/b_i^{\rm{space}} ) =0$ (S/B threshold equal to 1), are signal-like events, while lines lying below it are background events. $\Delta t_j$ are the time windows of consecutive doublets.}
\vspace{-0.2cm}
\label{fig:geotimeDTspace}
\end{figure*}

In the first step, we select flare candidates from the data using solely space information. Signal-like events are identified based on the ratio of their spatial signal PDF to the spatial part of the background PDF, which has to fulfill the following condition: 
\begin{equation}
s_i^{\rm{space}}/b_i^{\rm{space}}> \rm{S/B} \label{sb-ratio} \rm{,}
\end{equation}
where index $i$ numbers our events and S/B is an adjustable threshold that can be tuned as needed.
Next we extract all doublets of consecutive signal-like events to identify all possible time windows $\Delta t_j$ that compose the flares contribution. 
This procedure is visualized in Figure \ref{fig:geotimeDTspace}, where events collected over the time period $\Delta T_{\rm{data}}$ are shown as colored vertical lines. 
Color indicates timing of events, with blue corresponding to earlier times and red to later times. Signal-like events are lines extending above black horizontal line defined by $\log ( s_i^{\rm{space}}/b_i^{\rm{space}} ) =0$ (in this example, we set $\rm{S/B}=1$). All lines lying below it are background events. Time intervals between consecutive signal-like events (i.e. consecutive doublet time windows) are denoted by $\Delta t_1$, $\Delta t_2$, $\Delta t_3$, and so on.

In the second step, for each signal-like doublet, we calculate its significance by maximizing test statistic. We use a standard test statistic from \cite{BRAUN2008299} (see equation \ref{bb})
with addition of a marginalization term 
to provide a more uniform exposure for finding doublets of different widths \citep{GORA2011201}. The resulting test statistic is given by
\begin{equation}
    \rm{TS}_{\Delta t_j}(n) = - 2 \log \left[ \frac{\Delta T_{\rm{data}}}{\Delta t_j} \frac{ \mathcal{L}(0) }{ \mathcal{L}(n)} \right] \rm{.}
    \label{marg}
\end{equation}
Note that by construction of the signal PDF $s_i$ (see equations \ref{llh_sig_time} and \ref{sig_time}) only events within time window of
a considered doublet $\Delta t_j$ are taken into account, thus in this step full space-time information is used. 

\begin{figure}[t]
\vspace{-0.9cm}
\begin{center}
\includegraphics[width=1\textwidth]{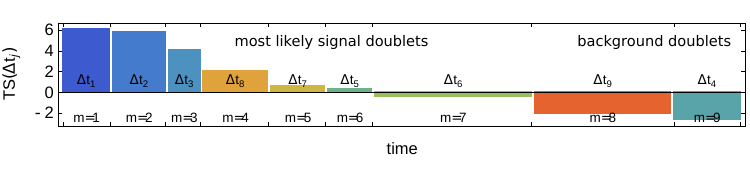}
\end{center}
\vspace{-0.9cm}
\caption{Doublets sorted according to their maximum value of the test statistics $\rm{TS}_{\Delta t_j}$, exemplary realization. Each colored box represents a single doublet. The width of the box is equal to the time window of the given doublet. The color indicates position of a doublet in time. Doublets are re-numbered according to their significance by introducing a multiplicity index $m$.
}
\label{fig:pij}
\vspace{-0.4cm}
\end{figure}

Next, all doublets are sorted and re-numbered (introducing multiplicity index m) according to their significance, which increases with the  maximum value of $\rm{TS}_{\Delta t_j}(n)$, which we denote by $\rm{TS}_{\Delta t_j}$.
This procedure is shown in Figure \ref{fig:pij}. Colored boxes are doublets sorted according to the maximum value of their test statistic. The doublet with multiplicity index $m=1$ is the most significant one. 
The width of the box is equal to the time window of the given doublet and color denotes doublet position in time. Doublets lying on the left side are most likely signal doublets, whereas those 
on the right side most likely are background doublets. 
Now, the main challenge lies in separating them. In other words, the goal is to determine
optimal $m$, i.e. $\rm{M}_{\rm{opt}}$, which effectively separates signal doublets from the background. Once we find $\rm{M}_{\rm{opt}}$, we can infer that flare or flares we are looking for most likely consists of $M_{\rm{opt}}$ doublets, ranging from $m=1$ up to $m=M_{\rm{opt}}$.
The time duration $\Delta T$ would be then the sum of the time windows corresponding to the most probable signal doublets.

\begin{figure}[t]
\begin{center}
\includegraphics[width=1\textwidth]{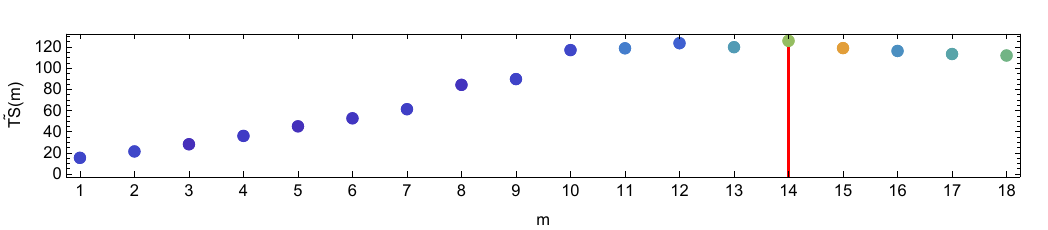}
\end{center}
\vspace{-0.5cm}
\caption{
Test statistic $\stackrel{\sim}{\rm{TS}}$ as a function of multiplicity
$m$, an exemplary realization for a single flare.
Each point represents the maximum possible value of the considered test statistic for a given multiplicity index $m$, i.e. assuming that the flare(s) consists of $m$ most significant doublets.
The color indicates position of multiplets in time. 
Here, the optimal multiplicity value, denoted by $M_{\rm{opt}}$, for which our test statistic reaches a maximum is $M_{\rm{opt}}=14$. Note the characteristic single-step shape, where the test statistic increases with $m$ and reaches a single plateau at large values of m. 
}
\vspace{-0.2cm}
\label{fig:mopt}
\end{figure}
The third and last step of our method is application of the stacking analysis to find $\rm{M}_{\rm{opt}}$. One-event signal probability $s_i$ is replaced by the weighted sum of signal sub-terms over m doublets, with weights $w_j = \rm{TS}_{\Delta t_j}$, i.e. we make the replacement
\begin{equation}
    s_i \rightarrow s^{\rm{tot}}_i (m) = \sum_{j=1}^m w_j s_i^j / \sum_{j=1}^m w_j \rm{,}
\end{equation}
where $s_i^j$ is the signal probability of the $i$th event calculated for the $j$th doublet ($s_i^j=0$ if the event is outside of the time window of the considered doublet). 
Then we use the standard likelihood and test statistic with stacking term $s_i^{\rm{tot}}(m)$, thus making following replacements: $\mathcal{L}(n) \rightarrow \mathcal{L}(n,m)$ and $\rm{TS}(n) \rightarrow \stackrel{\sim}{\rm{TS}}\hspace{-0.1cm}(n,m) = - 2 \log \left[ \mathcal{L}(0)/\mathcal{L}(n,m) \right]$.
Finally, we maximize $\stackrel{\sim}{\rm{TS}}\hspace{-0.1cm}(n,m)$ to find the optimal value of multiplicity of the flare $\rm{M}_{\rm{opt}}$, for which our test statistic reaches a maximum. 
This procedure is illustrated in Figure \ref{fig:mopt} for the case of a single flare.
$\rm{M}_{\rm{opt}}$ is the number of the most significant, thus not necessarily consecutive doublets, which makes the method sensitive for multiple flares.
As mentioned above, $\rm{M}_{\rm{opt}}$ determines the total duration of the flare(s), given by $\Delta T_{\rm{flare}}= \sum_{m=1}^{M_{\rm{opt}}} \Delta T_m $.
Similarly to the standard method, the estimator $n_s$ for the number of signal events is obtained by maximizing the final test statistic $\stackrel{\sim}{\rm{TS}}\hspace{-0.08cm}(n,\rm{M}_{\rm{opt}})$.

\begin{figure}[t]
\begin{center}
\vspace{-0.5cm}
\includegraphics[width=0.38\textwidth]{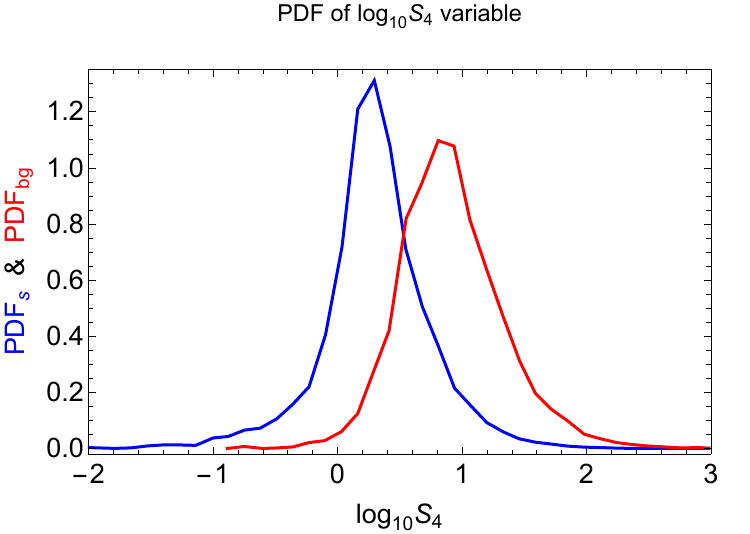}
\includegraphics[width=0.58\textwidth]{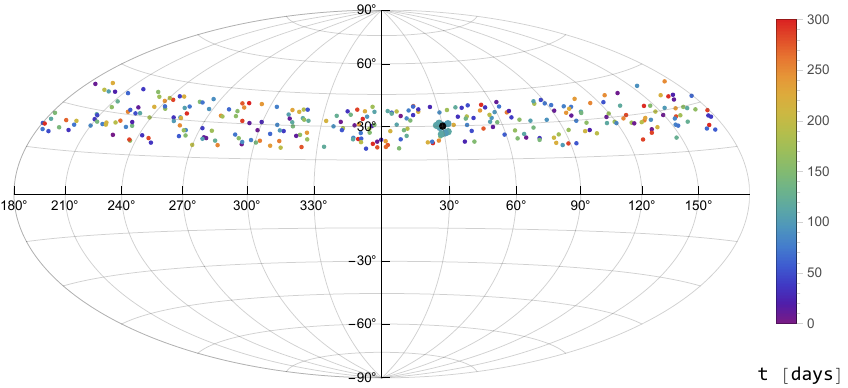}
\end{center}
\vspace{-0.5cm}
\caption{(\textbf{Left}) PDFs of $\log_{10} S_4$, obtained from simulations for photon-initiated (signal, blue) and proton-initiated (background, red) showers (Auger Collab., \citeyear{universe8110579}). (\textbf{Right}) An example of scrambled map shown for illustrative purposes. To
speed up the computation, instead of considering the entire sky, we only take
into account a small declination band that is wide enough to accommodate
all signal events. The source and background events are shown 
as a black dot and colored dots, respectively. The source is surrounded
by bigger dots, which are signal events. The color indicates the time of
events. 
}
\vspace{-0.4cm}
\label{fig:geo1}
\end{figure}

\section{$S_b$ photon tag}

The $S_b$ variable is commonly used to discriminate between photon-initiated and hadronic
showers. It can be defined as $S_b = \sum_k S_k \left(R_k /1000 \hspace{0.1cm} \rm{m} \right)^b$ 
\citep[Auger Collab., 2022]{ROS2011140}, where the summation is over all detectors with non-zero signals $S_k$, and $R_k$ is the distance of the $k$-th detector from the shower axis.
The b parameter can be chosen somewhat freely, however, for Auger photon searches, typically $b=4$ is used. Here we adopt the same value.
To enhance the sensitivity of the stacking method for photon search, we have applied a photon tag, which consists of a set of probability density functions (PDFs) that classify each event as more likely to be either a photon or background. As the photon tag, we use PDFs of $\log_{10} S_4$, obtained from simulations for photon-initiated (signal, $\rm{PDF}_s$) and proton-initiated (background, $\rm{PDF}_{bg}$) showers (Auger Collab., \citeyear{universe8110579}), see the left panel of Figure \ref{fig:geo1}. 
 To apply the $S_4$ photon tag, 
 we make the following replacements: $s_i^{\rm{space}} \rightarrow s_i^{\rm{space}} \rm{PDF}_s(S^i_{4})$ and $b_i^{\rm{space}} \rightarrow b_i^{\rm{space}} \rm{PDF}_{bg}(S^i_{4})$, where $S^i_{4}$ is the value of the $S_4$ variable for the $i$-th event.

\section{Monte Carlo test}

To perform Monte Carlo test of the stacking method, we randomly generated many scrambled sky maps containing both signal and background events. In each map, the background events were uniformly distributed over the sky and over the entire data collection period $\Delta T_{\rm{data}}$. 
The signal events follow a spatial Gaussian distribution centered around the source and are uniformly distributed over the
flare or flares duration $\Delta T_{\rm{flare}}$. 
Additionally, the start time of each flare is randomized.
An example of such a scrambled map is shown in the right panel of Figure \ref{fig:geo1}. 
For each map, 
we obtain estimator $n_s$ of the number of injected signal events $N_s$, estimator $\Delta T$ of the flare or flares duration $\Delta T_{\rm{flare}}$ 
and the maximum value of the test statistic $\stackrel{\sim}{\rm{TS}}\hspace{-0.1cm}(\rm{M}_{\rm{opt}})$, which can be used to draw conclusion about the statistical significance of the obtained result.

\begin{figure*}[t]
\vspace{-0.7cm} 
\begin{center} 
\includegraphics[height=4.15cm,trim={0cm 0cm 0cm 8cm},clip]{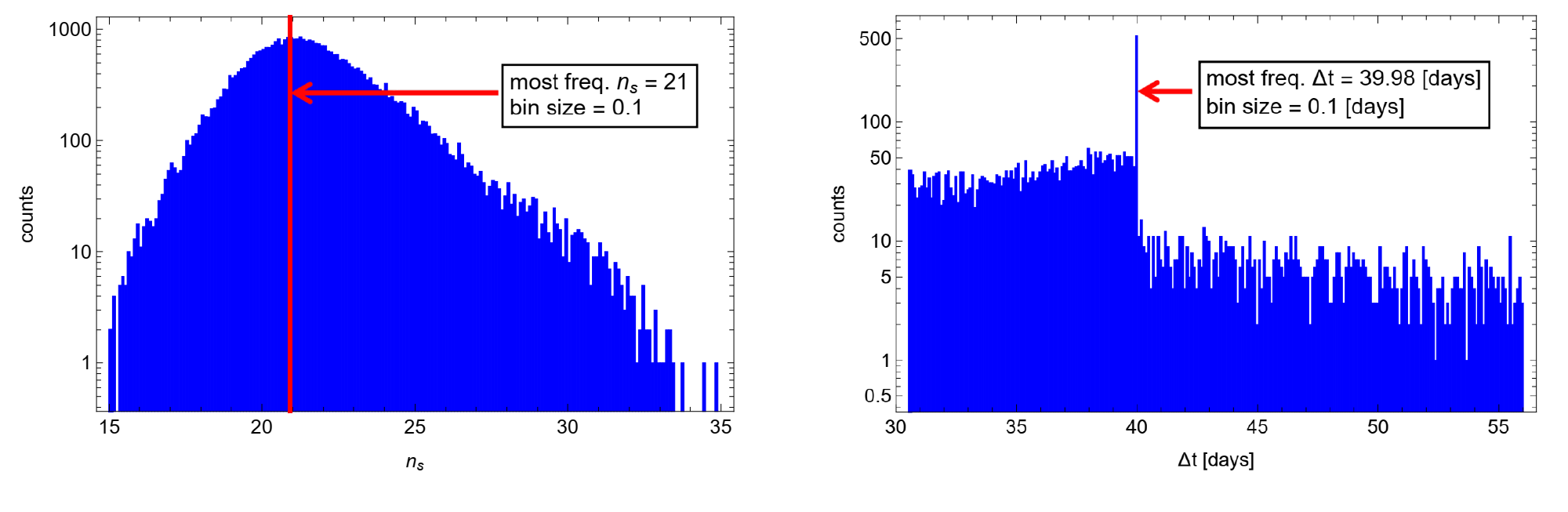}
\vspace{-1.0cm} 
\caption{Distributions of $n_s$ (\textbf{Left}) and $\Delta T$ (\textbf{Right}) obtained for the Monte Carlo test with triple flare (durations of 10, 10, and 20 days) with $N_s = 20$ signal events. The most frequent value of the given distribution, 
corresponding to the location of the highest peak, is presented in a box on individual plot. The number of injected signal events $N_s = 20$ and total flares duration $\Delta T_{\rm{flare}} = 40$ days are well recovered. The higher number of counts in the $\Delta T$ distribution at bins with $\Delta T < 40$ days indicates that the stacking method remains more sensitive to shorter flares, despite the inclusion of a marginalization term  $- 2 \log \left(\Delta T_{\rm{data}} / \Delta t_j \right)$ in the test statistic (see equation \ref{marg}).  
}
\vspace{-0.8cm}
\label{results}
\end{center}
\end{figure*}

We performed Monte Carlo simulations to assess the effectiveness of the stacking method in detecting flares and accurately retrieving the number of signal events and the flares duration. Figure \ref{results} shows exemplary distributions of the estimators $n_s$ and $\Delta T$, obtained for the case of three flares with durations 20, 10, and 10 days, and a total number of signal events of 20. The most frequent value in each distribution, corresponding to the location of the highest peak, 
is indicated in a box on the individual plot. These values are regarded as the best estimates of the parameters we aim to determine. As shown, the number of signal events and the total flare duration of 40 days are well recovered.

\section{Discovery potential}

Discovery potential tells us how many signal events are needed
to claim discovery of a cluster of events in data. It can be used
to compare different methods. By definition, the discovery threshold is the number of signal events required to achieve a p-value less than $2.87 \times 10^{-7}$ (one-sided 5$\sigma$) in 50$\%$ of the maps. To calculate this threshold we first use many scrambled sky maps to obtain the probability distribution of the global test statistic $\stackrel{\sim}{\rm{TS}}\hspace{-0.08cm}(\rm{M}_{\rm{opt}})$ for background-only simulations. We then fit an exponential curve to the tail of this distribution to determine the value of the test statistic corresponding to the threshold for the 5$\sigma$ excess.
Subsequently, we analyze the distributions of $\stackrel{\sim}{\rm{TS}}\hspace{-0.1cm}(\rm{M}_{\rm{opt}})$ for scenarios with varying numbers of signal events injected into the background. The number of injected signal events for which the median of the test statistic distribution reaches the 5$\sigma$ threshold for the background is our discovery threshold, i.e. the minimum number of signal events needed for discovery of flare(s). Note that this value must be calculated separately for single and multiple flares of different duration.

\begin{figure*}[t!]
\vspace{-0.9cm} 
\begin{center} 
\includegraphics[width=0.95\textwidth,angle=0]{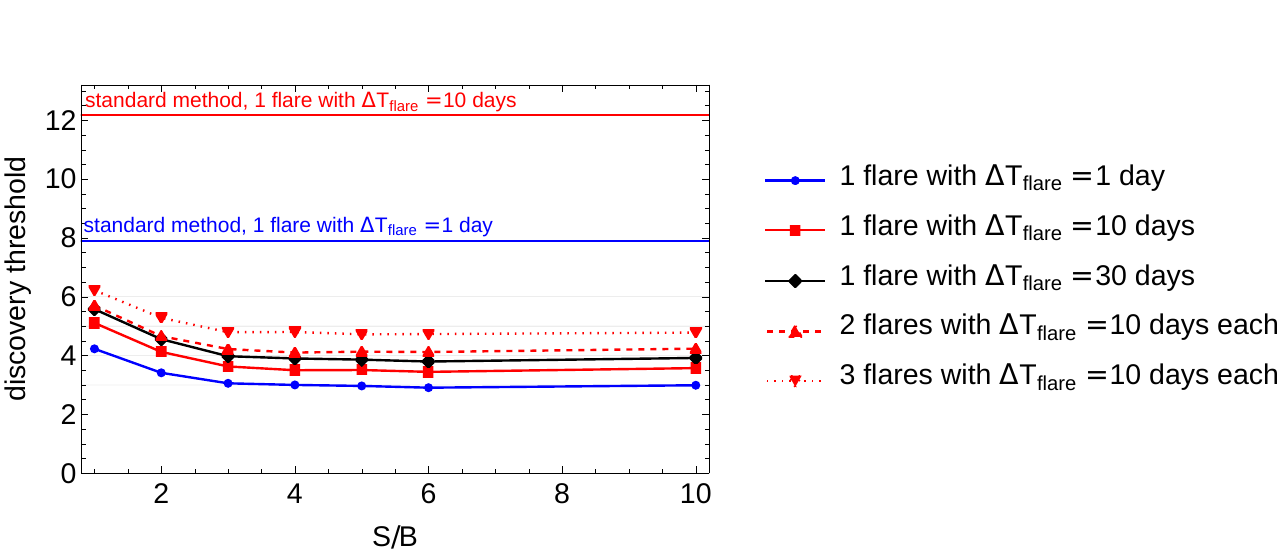}
\end{center}
\vspace{-0.7cm} 
\caption{Discovery thresholds for the stacking method with the $S_4$ photon tag are presented as a function of the S/B threshold for signal-like events. Shown are the results for single and multiple flares of different duration. For higher S/B thresholds, only a few events are required to detect flare(s). For comparison, the discovery thresholds for the standard unbinned likelihood method without a photon tag by \citet{BRAUN2008299}, are also shown. In this case, no pre-selection S/B threshold is applied, and results are provided for a single flare with duration of 1 day (blue line) and 10 days (red line). All the results have been obtained for $\Delta T_{\rm{data}} = 3150$ days and 595 of background events.}
\label{disco}
\vspace{-0.3cm}
\end{figure*}


Discovery thresholds for the stacking method with the $S_4$ photon tag, as a function of the threshold for signal-like events S/B, are shown in Figure \ref{disco}.
For higher S/B thresholds, only a few events are needed to detect flare(s). In contrast, the standard unbinned likelihood method without photon tag \citep{BRAUN2008299} requires more signal events, particularly for longer flares.

\section{Summary}

We proposed an improved method, referred to as the stacking method, to search for space-time clustering in UHECR (air-shower) data,
potentially providing evidence of ultra-high-energy (UHE) photons that may originate from astrophysical flares. It can also be used to improve limits on UHE photon flux and facilitate the search for sources of neutral UHECRs.

The stacking method offers several advantages:
it is faster than the standard method, more sensitive to weak flares of any shapes and capable of detecting multiple flares. Additionally, it can accurately recover the number of signal events and the duration of flare(s). Remarkably, the stacking method with the $S_4$ photon tag requires only a few events to discover photon flare(s)!

\small{Acknowledgments: 
The authors would like to thank the colleagues from the Pierre Auger Collaboration for all the fruitful discussions. 
We want to acknowledge the support in Poland from the National Science Centre, grants No. 2020/39/B/ST9 /01398 and 2022/45/B/ST9/02163 as well as from the  Ministry of Science and Higher Education, grant No. 2022/WK/12.
}

\bibliographystyle{apalike}
\bibliography{sample}

\end{document}